\documentclass[12pt,tightenlines,
    onecolumn,
    preprintnumbers,
    amsmath,
    amssymb,
    prd,
    showpacs
]{revtex4}
\usepackage{epsfig,bbm,
}
\usepackage{graphicx}%
\usepackage{amsmath}
\usepackage{amssymb}
\usepackage{bm}
\usepackage{enumerate}
\usepackage{colordvi}
\input colordvi
\usepackage{color}

\newcommand{\nn}{\nonumber}

\begin{document}
\title{Black hole as an Information Eraser}

\author{Hyeong-Chan Kim}
\email{hyeongchan@sogang.ac.kr}
\affiliation{Center for Quantum Spacetime, Sogang University,
Seoul 121-742, Republic of Korea
}%
\author{Jae-Weon Lee}
\email{jjlee@daejin.ac.kr}
\affiliation{School of Computational Sciences, Korea Institute for Advanced Study, 207-43 Cheongnyangni 2-dong, Dongdaemun-gu, Seoul 130-012, Korea
}%
\author{Jungjai Lee}
\email{jjlee@daejin.ac.kr}
\affiliation{ Department of Physics, Daejin University,
Pocheon, 487-711, Korea.
}%
\date{\today}%
\bigskip

\begin{abstract}
\bigskip
We discuss the identity of black hole entropy and show that the first law of black hole thermodynamics, in the case of a Schwarzschild black hole, can be derived from Landauer's principle by assuming that the black hole is one of the most efficient information erasers in systems of a given temperature. The term ``most efficient" implies that minimal energy is required to erase a given amount of information.
We calculate the discrete mass spectra and the entropy of a Schwarzschild black hole assuming that the black hole processes information in unit of bits.
The black hole entropy acquires a sub-leading contribution proportional to the logarithm of its mass-squared in addition to the usual mass-squared term without an artificial cutoff.
We also argue that the minimum of the black hole mass is $\sqrt{\log 2/(8\pi)}M_P$.
\end{abstract}
\pacs{04.70.-s}
\keywords{black hole, Landauer's principle, quantum black hole}
\maketitle

\section{Introduction}
The close connection between quantum information theory and general relativity have been discussed by Peres and Terno~\cite{Peres} and Hosoya~\cite{Hosoya}.
One of their results is that the generalized second law of black hole thermodynamics is satisfied even with a quantum version of a gedanken experiment with a detector falling into the black hole~\cite{Hosoya}, which was reexamined in Ref.~\cite{Song}.
Recently, the role of vacuum entanglement around a black hole horizon has been attracting much attention~\cite{Brustein,Srednicki,Plenio2,Bombelli,Terashima,Mukohyama,Ryu,Fursaev,Solodukhin}.
Due to its area proportionality, entanglement entropy~\cite{Srednicki,Plenio2} appears to be a natural candidate for the black hole entropy~\cite{Bombelli,Terashima,Mukohyama}.
Especially, Ryu and Takayanagi~\cite{Ryu} and Fursaev~\cite{Fursaev} studied the entanglement entropy in the context of holographic anti-de Sitter/conformal-field-theory correspondence and Solodukhin~\cite{Solodukhin} generalized it to include the entanglement entropy of black holes living on the boundary of anti-de Sitter space.
The role of information theory on cosmology was also discussed~\cite{lee} in relation to  {the} dark energy from information erasure.
Since the entanglement entropy is related to non-locality of quantum theory, quantum information theory~\cite{Landauer,Plenio} can play an important role in black hole physics.
The relation of the entanglement to Hawking radiation was also studied~\cite{Ahn,Braunstein}.
In Ref.~\cite{Fursaev}, it is conjectured, on the basis of the relation between the entropy and action, that in a fundamental theory the ground state entanglement entropy per unit area equals $k_B c^3/(4G\hbar)$. According to their argument, the black hole entropy might be due to the vacuum entanglement.
Horowitz and Maldacena~\cite{Horowitz} proposed a conjecture on the final state of black hole evaporation as a completely entangled state, in which the black hole  {evolves unitarily}.
It is also noted that information inside the horizon can be transferred to the outside by means of quantum teleportation~\cite{Lloyd,Ge}.

In this paper, we study two new aspects of blackhole physics.
First, we show that the first law of blackhole thermodynamics can be derived from Landauer's principle~\cite{Landauer,Plenio} of information theory.
Second, the information erasing process related to Landauer's principle happens at the just outside of the blackhole horizon.
This result give new insights on the information loss paradox of the blackhole physics.
Since Landauer's principle is based on the second law of thermodynamics, this derivation not only provide a close link between the information theory and blackhole physics but also relates the second law to its first law.
A black hole satisfies the four laws of black hole mechanics~\cite{Bardeen}, which are essentially the same as those of thermodynamics.
Especially for a Schwarzschild black hole, the first law
\begin{eqnarray} \label{eq:1stlaw}
d(M c^2)= \frac{\kappa c^2}{8\pi G}\,d A
\end{eqnarray}
relates its mass $M$ to its horizon area $A$, where $\kappa$ is the surface gravity of the black hole.
Hawking~\cite{Hawking} found that a black hole really radiates as if it is a black body with temperature,
$
T_H= \frac{\hbar}{k_B c}\frac{\kappa}{2\pi},
$
by studying quantum field theory in a classical background black hole spacetime.
From the first law~(\ref{eq:1stlaw}) and the analogy of $Mc^2$ with thermodynamic energy, the entropy of the black hole~\cite{Bekenstein} is given by
\begin{equation}\label{S:A}
S_{bh}= \frac{k_B}{4}\frac{A}{l_p^2},
\end{equation}
where $l_p =\sqrt{\frac{\hbar G}{c^3}}$ is the Planck length. The explicit presence of $\hbar$ in the Hawking temperature denotes the quantum mechanical nature of the black hole radiation.

The first law of black hole thermodynamics relates the mass to the entropy of a black hole, however the origin of this link is unknown.
There have appeared many different approaches that have aimed at  explanation of the origin of the black hole entropy including string theory~\cite{string} and the brick wall method~\cite{brickwall,mhlee}.
The quantum mechanical nature and the geometric property of black hole entropy give rise to many speculations on their quantum gravity origin, such as the entropy bound~\cite{Bekenstein2,Bousso}, the holographic principle~\cite{tHooft,Susskind}, the holographic dark energy model~\cite{Li}, and the spacetime noncommutativity~\cite{Doplicher,kappa}.
These approaches try to find an appropriate model of the horizon or internal structures of the black hole, which provide the correct description of black hole entropy.

In this paper, we suggest an alternative way to explain the relation between the black hole entropy and its mass.
Rather than guessing at the internal structure of the black hole, we  investigate ``What happens when information is lost at the horizon?" employing Landauer's principle.
A merit of this approach is that we do not need to worry about ignorance of the internal structure of the black hole.
We argue that the black hole mass and its relation to the entropy come from  the postulation of black hole information erasure: {\it  A black hole is one of the most efficient information eraser in systems of a given temperature.}
This postulation arises from the fact that ``The black hole is one of the maximal entropy object for a given global quantities such as mass, angular momentum, and charge".
We also show that the black hole mass becomes discrete if the lost information is dealt with in unit of bits.
These results would be general, since we directly resort to {\it our ignorance on the lost information.}

In this work, we describe the
Schwarzschild black hole using the information erasing procedure in Sec. II.
Some puzzles about black hole entropy are discussed and resolved using Landauer's principle.
In Sec.~III, the
discrete black hole mass spectrum is obtained by using the fact that the information is quantized in unit of bits. It is
  {argued} that there is a minimum value of black hole mass.
In section~IV,
we summarize and discuss the results. We also briefly mention the possibility of experiments that could probe the minimum mass of black holes.

\section{Black hole as an information eraser} \label{sec:class}
In this section, we discuss information erasing process which happens at the just outside of the blackhole horizon in comparison with a typical information erasing process by using an atom and a cylinder.

\subsection{Landauer's principle}

~Let us briefly review Landauer's principle.
In 1961, Rolf Landauer had the important insight that there is a fundamental asymmetry in the way that the nature allows us to process information~\cite{Landauer}. There, he showed that copying classical information can be done reversibly and without wasting energy, but when information is erased there is always an energy cost larger than $k_B T \log 2$ per classical bit to be paid~\cite{Plenio}.
 Recently, this principle has been challenged in Refs.~\cite{Land:challenge} but was depended in Refs.~\cite{Land:protect,PhysRevA.61.062314}.
\begin{figure}[htb]
  \begin{center}
  \includegraphics[width=.3\linewidth]{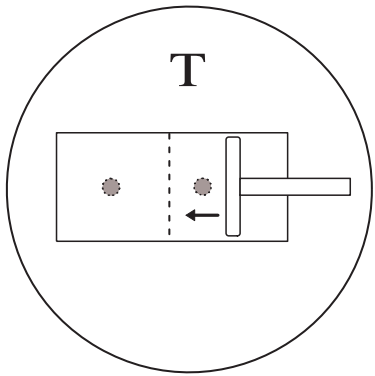}\\
  \end{center}
  \caption{Information erasing process of a 1-bit memory. The memory consists of a cylinder and an atom. }\label{fig:info}
\end{figure}
Explicitly, consider a 1-bit memory which consists of a cylinder and an atom in it as shown in Fig.~\ref{fig:info}.
A thermal bath with temperature $T$ keeps in contact with the cylinder.
The location of the atom, at the left or the right partition, represents a memory.
To erase this memory we use the piston to push the atom into the left partition regardless of its initial position.
Then, the initial information of the atom is erased irreversibly because we will never find out where the atom was originally.
According to Landauer's principle the entropy of the cylinder {\it decreases} by $k_B\log 2$ at the cost of at least $k_BT\log 2$ free energy consumption.
This energy is eventually converted to thermal energy and increases the total energy of the whole system (the bath+the cylinder) by
\begin{eqnarray} \label{eq:dE}
\delta E \geq k_B T \log 2 .
\end{eqnarray}

\subsection{Information erasing around a Schwarzschild blackhole  }

Let us briefly introduce two coordinate-dependent phenomena, Hawking radiation and the ``information freezing".
The quantum mechanical nature of a black hole is represented by the Hawking radiation, which was explained as a quantum tunneling effect around the horizon~\cite{Parikh}.
Even though it is physical, Hawking radiation is dependent on the choice of coordinates in the sense that a freely falling observer can not feel its presence since the system is in his vacuum state, the Hartle-Hawking vacuum.
This
 {coordinate dependence} is explained due to the presence of the event horizon and the coordinate singularity there.
The concept of ``information freezing" appears when we consider a black hole spacetime in which a freely falling rocket periodically emits
an outgoing light ray of a given frequency.
An observer freely falling with the rocket may notice that the rocket simply enters into the black hole and may not even appreciate the presence of the horizon.
On the other hand, from the viewpoint of a stationary observer outside,
it takes infinite time for the rocket
to get to the horizon and the waves radiated by the rocket experience gravitational red-shift.
In this sense, the information
contained in the rocket is frozen at the horizon.
To emphasize this phenomenon, the black hole was once called as a frozen star.

Hawking radiation is regarded as a physical phenomenon and it decreases the mass of the black hole.
In contrast, the information freezing at the horizon is regarded as a simple coordinate artifact and is short of physical implication.
Even though there exist differences between information freezing and Hawking radiation, their origins are related:
The reference frame of the observer determines the observable radiation and accessible information.
Therefore, it is plausible to think that the information freezing may also
have physical implications.
%
In this paper, we provide a possible interpretation, which resolves the puzzles on the relation between the mass and the entropy of a black hole.  We also discuss the physical meaning of the information freezing and erasing which happen at the black hole horizon.
\begin{figure}[htb]
    \begin{center}
  \includegraphics[width=.4\linewidth]{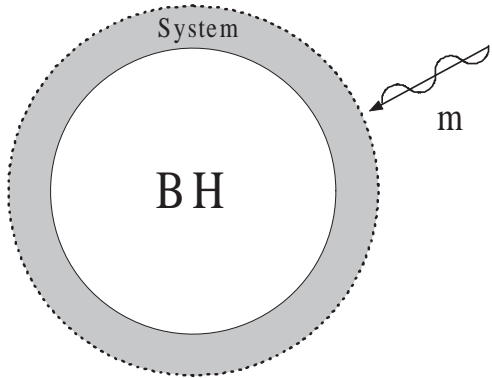}
  \end{center}
  \caption{The black hole is surrounded by a ``system", the intermediate region of information. The energy going into the black hole is used to erase the information of the system. }\label{bh}
\end{figure}

To resolve the entropy puzzles, we suggest that the frozen information composes a ``system" in Fig.~\ref{bh} around the horizon.
We assume that the energy composing the information of the ``system" is finite at the horizon. Therefore,  the energy vanishes from the viewpoint of an asymptotic observer because of the redshift factor.
The ``system" plays the same role as the cylinder in the previous example.
The ``system" is in thermal contact with the black hole, which plays the role of thermal bath with the Hawking temperature $T=T_H=\frac{\hbar c^3}{8\pi k_B GM}$.
For example, in systems with entangled states, the use of the Hawking temperature as the system's temperature was justified~\cite{Ohta}.
An in-falling particle interacts with the ``system" not with the black hole directly. The energy of the in-falling particle is used to erase the information of the ``system" and is transferred to the black hole through the thermal contact between the ``system" and the black hole.
Therefore, the energy of in-falling particle corresponds to the work done by the piston in Fig.~1.
In this sense, we consider the thermodynamics of the ``system"  rather than that of the black hole itself. In this way, we avoid the difficulty in dealing with the one bit of information in black hole physics.

Note that the information erasing mentioned in this paper is much different from the information hiding such as putting some information in a strongbox.
The hidden information may be uncovered by opening the strongbox.
However, the erased information cannot be recovered by any mean.
In addition, the information erasing process occurs at just the outside of the horizon.
In this sense, we may conclude that only the energy enter into the blackhole horizon and the information does not.
This can be a resolution of the information loss paradox.
Since the information does not enter into the horizon but erased outside the horizon, no information may enter into the blackhole horizon.
In this sense, no information is lost from the point of view of the outside observer.

For simplicity, we assume that the information in the ``system" consists of classical bits.
In general there are other possibilities: The information may not be classical or the information would consists of trits, etc.
We divide the in-falling procedure of a particle into three parts:
First, the energy of the particle is used to erase the information of the ``system" .
During the information erasing process of one bit, the entropy of the one bit ``system" decreases to zero.
This entropy is transferred to the thermal bath through the thermal contact and increases
the black hole entropy by $k_B\log 2$ per bit.
The energy provided to the ``system" is also absorbed by the black hole.
Second,
the black hole horizon increases as the entropy and engulfs the system itself.
Third, the information of the in-falling particle is frozen at the horizon and eventually composes the {\it new} ``system" itself.
This procedure explains how the ``system" is constructed and how the incoming information is erased consistently.

The energy used to erase the system's information satisfies Landauer's principle.
Here we postulate a principle of information erasure for a black hole: ``{\it A black hole is one of the most efficient information erasers in systems of a given temperature.}"
This principle is in parallel with the fact that the black hole is a maximal entropy object for a given set of global quantities in the sense that minimum energy is required to erase a given quantity of information.
Physically, this implies that the information erasing process is optimal so that the energy provided to the ``system" by the in-falling matter during the erasing process {\it saturates the Landauer's bound}, $\delta E= k_B T_H \log 2$.
Note that this is a classical information erasing process of one bit and we do not need any other assumption except for Landauer's principle.
As a result, this saturating energy $\delta E$ is transferred to the black hole, {since the ``system" is in thermal contact with the black hole which plays the role of the thermal bath and increases} its mass by
\begin{eqnarray} \label{eq:dM}
\delta (Mc^2)=\delta E= k_B T_H \log 2.
\end{eqnarray}
The right hand side of Eq.~(\ref{eq:dM}) is just the black hole temperature $T_H$ times the increased entropy, $\delta S= k_B \log 2$, of the black hole  {with the} unit of $k_B$.
Note that the relation~(\ref{eq:dM}) does not come from the first law of black hole thermodynamics.
It just says that the energy $k_B T_H \log 2$ {is transferred
from the infalling particle to the ``system" erasing one bit of} information in the ``system"  {and finally}
 to the black hole through the thermal contact.
In this way, the increase of the black hole mass $\delta M$ is related {to} 
 the information erasing of the ``system" directly.

\section{Discrete black hole mass from information erasing }
From this point on in this paper, we will use the natural units $k_B=1=G$. We consider  {a Schwarzschild black hole mass spectrum} by using a sequence of $(N-1)$-bits of information-erasing process bit by bit.
 { Therefore, we restrict ourselves only to the uncharged spherically symmetric cases.
Consider  a one bit ``system" of information interacting with a small black hole with entropy $S_1=\log 2$. We leave the black hole mass $M_1$ as a free parameter.}
We determine $M_1$ later so that its value maximizes the entropy of macroscopic black holes.
To avoid the conical singularity of the Wick rotated black hole metric, it is natural to select the temperature of the ``system" to be the black hole temperature $T_1=1/(8\pi M_1)$.
Later in this paper, we discuss the possibility that the quantum gravity effect may change this temperature-mass relation.
After one bit of information erasing of the ``system", the black hole mass is increased by $M_2-M_1=\log 2/(8\pi M_1)$ from Eq.~(\ref{eq:dM}) and its temperature becomes $T_2=1/(8\pi)[M_1+ \log 2/(8\pi M_1)]^{-1}$, which becomes the new temperature of the ``system".
We may repeat this procedure which is represented by the recurrence formula
\begin{eqnarray} \label{eq:Mquant}
X_{n+1} &=& X_n+X_n^{-1},
\end{eqnarray}
where the temperature and the mass of the black hole are
\begin{eqnarray}\label{MT:X}
 \quad M_n= \sqrt{\frac{\log 2}{8\pi}}X_n,
    \quad T_n = \sqrt{\frac{1}{8\pi\log 2}}\frac{1}{X_n}, \quad n\geq 1,  \label{M:X}
\end{eqnarray}
respectively.
After $(N-1)$-repeated erasing processes, the informational entropy of the final black hole of mass $M=\sqrt{\frac{\log 2}{8\pi}}X_N$ is
\begin{eqnarray} \label{eq:Sn}
S_{bh}\equiv S_N=S_1+ (N-1) \log 2=N\log 2.
\end{eqnarray}
Note that the entropy is proportional to the number of bits $N$.
In the large $N$ limit, we have the limiting behavior: $\displaystyle \lim_{N\to \infty} \frac{X_N}{ \sqrt{2 N}}=1$.

\begin{figure}[htb]
  \begin{center}
  \includegraphics[width=.5\linewidth]{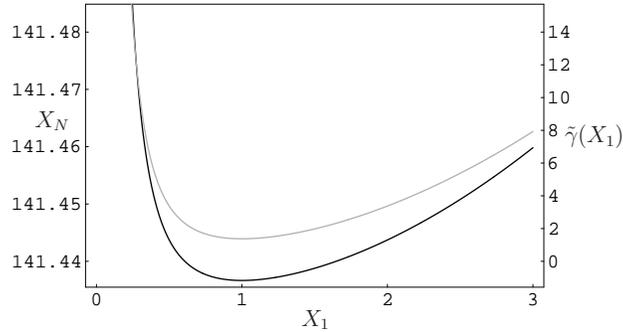}
  \end{center}
  \caption{$X_N(X_1)$ (black curve) and $\tilde \gamma(X_1)$ (gray curve) as functions of $X_1$.
  In this figure we use $N=10^4$. The minima of $X_N$ and $\tilde \gamma(X_1)$ are placed at $X_1=1$ and its value is $X_N \simeq 141.436659$ and $\tilde \gamma(1)\simeq 0.376569$. }\label{xn1}
\end{figure}

Instead of solving the recurrence formula~(\ref{eq:Mquant})
 {directly}, 
we try to find an approximate solution.
Summing over $n$ after multiplying $X_n$ to  {both sides of } Eq.~(\ref{eq:Mquant}), we get
\begin{eqnarray} \label{eq:N-sumX}
N-1=\sum_{n=1}^{N-1} X_n \delta X_n &=& \int_{X_1}^{X_N} X d X
    -\frac{1}{2}\sum_{n=1}^{N-1} \delta X_n^2 \\
    &=&  \frac{X_N^2-X_1^2}{2}-\frac{1}{2}\sum_{n=1}^{N-1} \frac{1}{X_n} \delta X_n \,,\nn
\end{eqnarray}
where $\delta X_n= X_{n+1}- X_n$.
To find an approximate expression for Eq.~(\ref{eq:N-sumX}), we introduce a large integer $H\gg N$ in Eq.~(\ref{eq:N-sumX}) and divide the summation into two parts: $1$ to $H$ and $H$ to $N$. Then, we approximate the second summation with an integral form $\int_{X_H}^{X_N} X^{-1} dX$ for large $N$ to get,
\begin{eqnarray} \label{eq:N:X11}
N-1
&=& \frac{1}{2}\left(X_N^2 -\log X_N -1- \tilde \gamma(X_1)\right) ,
\end{eqnarray}
where the function $\tilde \gamma(X_1)$ has a well defined value,
\begin{eqnarray}
\tilde\gamma(X_1) &\equiv&
    \lim_{H\rightarrow\infty}\left(
    \sum_{n=1}^{H-1}\frac{1}{X_n^2} -\log{X_H}\right)+X_1^2-1.  \label{eq:gamma:X1}
\end{eqnarray}
Note that $\tilde\gamma(X_1)$ is guaranteed to converge since the large $H$ limit of the term inside the parenthesis shows the same behavior as that of the definition of the Euler-Mascheroni constant because $X_n\sim\sqrt{2n}$ for $n\gg 1$.
$\tilde \gamma(X_1)$ is a function of $X_1$ only and is minimized at $X_1=1$ as seen in Fig.~\ref{xn1}.
This $X_1=1$ is not numerical but {\it exact}, since $X_n$ with $n\geq 2$ is a function of $X_1+X_1^{-1}$, of which the derivative vanishes at $X_1=1$.
As seen in Fig.~1, for fixed $N$, $X_N$ is minimized at $X_1=1$ too.

The entropy of a black hole is given as a function of $X_1$ and $X_N$ by
\begin{eqnarray*}
S_{bh}=\frac{\log 2}{2}\left[X_N^2-\log X_N+1
    - \tilde \gamma(X_1)\right] \,.
\end{eqnarray*}
For a given mass $M$, this entropy will be maximized for $X_1=1$, which determines the initial mass
\begin{eqnarray} \label{eq:M1}
M_1=\sqrt{\frac{\log2}{ 8\pi}}.
\end{eqnarray}
Therefore, the informational entropy of a black hole with mass $M=M_1 X_N$ is
\begin{eqnarray} \label{eq:S-X}
S_{bh}=  \frac{\log 2}{2}\left(\frac{M^2}{M_1^2}
    -\log \left(\frac{ M}{M_1}\right)
    +1-\tilde\gamma(1)\right),
\end{eqnarray}
where the mass is measured in Planck units and $\tilde\gamma(1)\approx 0.37657 $.

\begin{figure}[htb]
  \begin{center}
  \includegraphics[width=.5\linewidth]{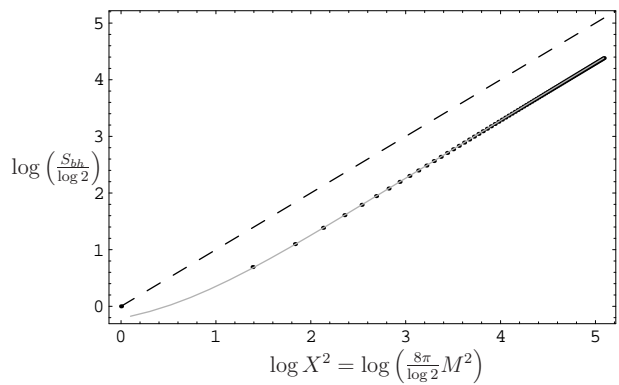}
  \end{center}
  \caption{ Entropy $S_{bh}/\log 2$ as a function of black hole mass squared $X^2=\frac{8\pi}{\log 2} M^2$ in log-log plot.
   The gray curve and the dashed line denote the logarithm of Eq.~(3.7) and the logarithm of the classical black hole entropy formula $4\pi M^2$, respectively. The dots denote the entropy for each discrete mass of black hole for $n=0,1,2 \cdots$ obtained from the recurrence relation~(3.1).
   The informational black hole entropy (dots) is slightly smaller than the classical expectation (dashed line). } \label{fig:Sbh}
\end{figure}

As seen in Fig.~\ref{fig:Sbh}, the approximate formula~(\ref{eq:S-X}) provides the correct black hole entropy for $n\geq 2$.
Because of the negative sign in front of $\log M$ in Eq.~(\ref{eq:S-X}), the informational black hole entropy is slightly smaller than the classical one given by the area law.
~ { For a macroscopic black hole with $M\gg M_1$, the logarithmic and constant terms in Eq.~(\ref{eq:M-N})}  {are} much smaller than $M^2/M_1^2$ term.  { Therefore, the relative difference between the classical result and the present formula becomes negligible for large $M$.}

In addition, the formula~(\ref{eq:S-X}) gives a discrete black hole mass because the informational entropy~(\ref{eq:Sn}) is $N\log 2$ with integer $N$.
The mass spectrum of the spherical black hole is given by solving,
\begin{eqnarray} \label{eq:M-N}
\frac{M^2}{M_1^2}-\log \left(\frac{M}{M_1} \right)=2N - (1-\tilde\gamma(1)) ,
\end{eqnarray}
where $N=1,2\cdots$ is an integer.

~ {In the large mass limit, ignoring the logarithmic term,} we arrive at the classical formula of the black hole mass and entropy:
\begin{eqnarray} \label{M:S:infor}
M(S)= \sqrt{\frac{S}{4\pi}} .
\end{eqnarray}
Note that the black hole mass is directly obtained from its entropy contents through the principle of information erasure for the black hole.
This implies that the whole mass of the Schwarzschild black hole comes from the thermal energy during the information erasing process of the ``system".
If some mass $m_0$ is allowed to enter  {behind} 
the horizon without information erasure, then the black hole mass formula should be changed to $M(S)=\sqrt{\frac{S}{4\pi}}+m_0$ due to the energy conservation law with $m_0$ being independent of the parameters of the black hole. However, the known formula for the entropy of the Schwarzschild black hole fixes $m_0=0$, which implies that no incoming energy goes behind 
the horizon without erasing the information of the ``system".
This is why the principle of information erasure for black holes
provides the correct relation between the black hole entropy and mass.

The quantization of black hole and its effects on  {the} black hole entropy were discussed by several authors~\cite{Bekenstein3,Mukhanov,Hod,Corichi}.
Based on the adiabatic invariance of the horizon area, Bekenstein~\cite{Bekenstein3} argued that the horizon area is quantized in {the} unit of $4 \log k$. In addition,
Hod~\cite{Hod} determined $k=3$ using the Bohr's correspondence principle and the asymptotic behavior of the ringing frequencies~\cite{Nollert} of  {the} black hole's quasi-normal modes.
Our result is slightly different from Hod's by the factor of $\log 2$.
If, for some reason, the black hole can not absorb the information with the unit of bits, or the information of the ``system" is  arranged not by bits but by trits (trinary digits), then the present result should be rewritten with $\log 3$. These results  approach those of Hod~\cite{Hod} and Corichi~\cite{Corichi} for large $N$. This is directly related to the quantum gravitational structure of the event horizon.

In the presence of quantum effects, the black hole horizon area may fluctuate, which makes the surface gravity and the temperature also fluctuate.
In addition, for small black holes, the change in geometry with  one bit of information erasure is non-negligible.
In this case, the assumption of the black hole as a heat bath with constant temperature fails. We present a rough estimation that can take care of these effects.
 {With the  {information erasure of} one bit},  {the temperature of the black hole} decreases to $T_{n+1}= \sqrt{\frac{1}{8\pi\log 2}} \frac{1}{X_n+X_n^{-1}}$ from $T_n$.
Therefore, to calculate the increased mass effectively, we
use an intermediate temperature $T_{n}'= \sqrt{\frac{1}{8\pi\log 2}} \frac{1}{X_n+\alpha_n X_n^{-1}}$ between $T_n$ and $T_{n+1}$, where $ \alpha_n$ is a $n$-dependent constant smaller than one.
This modification changes the recurrence relation to $X_{n+1}= X_n+ (X_n+\alpha_n X_n^{-1})^{-1}$ and alters the sub-leading logarithmic contribution of the entropy from $-\frac{\log 2}{ 2}\log M$ to $\frac{\log 2}{ 2}(\alpha -1)\log M$ if $\alpha_n=\alpha$ is independent of $n$.

\section{Summary and discussions }

We have postulated the principle of information erasure for a black hole: ``A black hole is one of the most efficient information erasers in systems of a given temperature".
Based on this postulation, we presented a direct explanation for the black hole mass from the contents of erased information by the black hole by using Landauer's principle.
This derivation of the first law of blackhole thermodynamics is based on the assumption that the infalling particle experiences a process of information erasure at just the outside of the blackhole horizon.
The information erasing process is much different from information hiding in the sense that it is irreversible.
If one want to recover the erased information, one need to know the erasing process explicitly.
Note also that the word ``most efficient information erasure" is different from the word ``maximal entropy".
The efficiency comes from the ratio of erased information for a given amount of energy and the ``maximal entropy" implies maximal entropy for a given volume of space.
The erasing process happens at an artificial ``system" of information located just outside of the horizon.
%
%
The present results illuminate the true nature of black hole entropy.
The content of missing information provides the correct relation 
between the entropy and the mass of a black hole.
Therefore, black hole entropy counts the content of missing information rather than the internal degrees of freedom of the black hole.
This also implies that all falling energies into the black hole are used to erase information around the horizon.
The details of the erasing process are an interesting subject for future research.
The connection between information theory and blackhole thermodynamics is widely under investigation recently~\cite{Peres,Hosoya,Song,Brustein,Maruyama}.
Therefore, the present paper will be useful in this respect.

We calculated the discrete black hole mass by using the fact that the information is quantized with the unit of bits.
We have shown that there exists a minimum mass of a black hole, which has the entropy corresponding to a 1-bit of information erasure.
The presence of the minimum of  the black hole mass $M_{\rm min}=M_1= \sqrt{\log 2/(8\pi)}$ also implies the  {existence} of the maximum of  {the} black hole temperature, $T_{\rm max}= (8\pi \log 2)^{-1/2}\simeq 0.24$.
Since  {the} Planck scale vacuum fluctuation may compose the Planck scale black hole, the existence of the maximal temperature of  {the} black hole implies the absence of black hole with temperature larger than the critical temperature $T>T_{\rm max}$, no matter how much energy is confined in a small region.
Another insight one gets from this result is that the black hole mass spectrum is discrete. Since the mass gap is big for a small black hole, a particle of small energy may have difficulties in entering the black hole horizon~\cite{Tye}.
This mass gap may prevent a small black hole from becoming a larger one and we need a mechanism to circumvent this situation.
If the Planck scale is about TeV as in  {the} brane world scenarios~\cite{brane}, the discreteness of the mini-black hole  {mass} might be observed at the Large Hadron Collider (LHC) in the near future.
If the center of mass energy of two colliding particles is high enough, a microscopic black hole can be produced.
The discreteness of the black hole mass spectrum could be observed from the peaks of the cross section graphs.



\begin{acknowledgments}
This work was supported by SRC Program of the KOSEF through the CQUEST grant R11-2005-021 (K), KRF-2008-314-C00063 (K,JJL), and the IT R\&D program of MIC/IITA (2005-Y-001-04, Development of next generation security technology; J-W.L).
\end{acknowledgments} 

\end{document}